\documentclass{article}
\usepackage{graphicx}
\usepackage{amsmath}

\input{tcilatex}

\begin{document}

\title{Regular Type III and Type N Approximate Solutions}
\author{Philip Downes, Paul MacAllevey, Bogdan Ni\c{t}\u{a}, Ivor Robinson.}
\maketitle

\begin{abstract}
New type III and type N approximate solutions which are regular in the
linear approximation are shown to exist. For that, we use complex
transformations on self-dual Robinson-Trautman metrics rather then the
classical approach. The regularity criterion is the boundedness and
vanishing at infinity of a scalar obtained by saturating the Bel-Robinson
tensor of the first approximation by a time-like vector which is constant
with respect to the zeroth approximation.
\end{abstract}

\section{Introduction}

Exact solutions of type III and type N built around twisting congruences are
difficult to obtain. Because of this, people have looked at approximate
solutions, in principal for the simpler, null case. In this paper we present
a method of obtaining a class of type III linear solutions based on
complexification of a non-twisting space-time and its complex transformation
into a twisting one.

Any real solution of a Lorentz invariant theory can be transformed, via a
complex Lorentz transformation, into a complex solution. When dealing with
linear solutions such as the results of a linear approximation to a
gravitational field, the real and imaginary parts are, in principle, new
solutions of the theory [1], [2]. When the additional requirement of
preserving the algebraic type of the original electromagnetic field or
gravitational field is considered, a complex Lorentz transformation is
insufficient. Recall that the algebraic type of the Weyl tensor is a
property of the self-dual and anti-self-dual parts taken separately.
Algebraic integrity is maintained between the real and complex solutions if
one proceeds by performing a complex Lorentz transformation on the self-dual
solution and then adding its complex conjugate to obtain the real part. This
keeps the algebraic structure of the Weyl tensor and in particular any null
bivectors which are coincident remain coincident. At this point note that
with the self dual and anti-self-dual solutions defining the principal null
congruence, the geometry of the principal null congruence has not been
preserved.

We carry out this procedure explicitly starting with a linear approximation
of a type III space-time built around the nontwisting Robinson congruence.
After complexifying and putting conditions for self-duality, we apply a
complex transformation to obtain the most general expression for type III
solutions built around what might be described as a generalized Kerr
congruence. A large sub-class of these solutions are shown to be regular
using a scalar obtained from their Bel-Robinson tensor.

This method has its limitations. First, one cannot extend it to obtain the
most general type III twisting solutions. Second, the obtained solution may
not be in a surveyable form .

\section{Self-dual Robinson-Trautman spaces}

We can complexify the Robinson-Trautman metric, in a purely formal way, by
taking the coordinates $\rho ,$ $\sigma ,$ $\zeta ,$ $\widetilde{\zeta }$ to
be independent complex variables and the defining functions $m(\sigma )$ and 
$p(\sigma ,\zeta ,\widetilde{\zeta })$ to be complex. It is then easy to see
that the Weyl tensor is self-dual if, and only if, $m=0$ and $p^{-1}p_{%
\widetilde{\zeta }\widetilde{\zeta }}$ is a function of $\widetilde{\zeta }$
only. By means of a coordinates transformation we can strengthen the last
condition to $p_{\widetilde{\zeta }\widetilde{\zeta }}=0.$ The line element
is then given by 
\begin{eqnarray}
ds^{2} &=&2d\rho d\sigma +\left( K-2\rho \frac{p_{\sigma }}{p}\right)
d\sigma ^{2}-\frac{2\rho ^{2}}{p^{2}}d\zeta d\widetilde{\zeta } \\
K &=&2(AB_{\zeta }-A_{\zeta }B),p=A+\widetilde{\zeta }B
\end{eqnarray}
where $A$ and $B$ are arbitrary functions of $\zeta $ and $\sigma .$ The
Weyl tensor is 
\begin{eqnarray}
C_{klmn} &=&\frac{p}{2\rho ^{2}}K_{\zeta }(M_{kl}N_{mn}+N_{kl}M_{mn})-\frac{1%
}{2\rho ^{2}}(p^{2}K_{\zeta })_{\zeta }N_{kl}N_{mn} \\
&&+\frac{p^{2}}{\rho }(p^{-1}p_{\zeta \zeta })_{\sigma }N_{kl}N_{mn}  \notag
\end{eqnarray}
where 
\begin{eqnarray}
N &=&\frac{\rho }{p}d\sigma \wedge d\zeta , \\
M &=&d\sigma \wedge d\rho -\frac{\rho ^{2}}{p^{2}}d\zeta \wedge d\overline{%
\zeta }.
\end{eqnarray}

\section{Change of coordinates}

It is worth remarking that, at least in a special case, one can put this
line element into the standard form for one with twisting rays. The special
case is defined by $p=A(\zeta )+\widetilde{\zeta }B(\zeta )$. The
transformation 
\begin{eqnarray}
\rho  &=&r-i\Sigma   \notag \\
\zeta  &=&z  \label{3} \\
\widetilde{\zeta } &=&\frac{r+ia}{r-ia}\widetilde{z}.  \notag
\end{eqnarray}
where $\Sigma =a\frac{A(z)-\widetilde{z}B(z)}{P}$ and $P=A(z)+\widetilde{z}%
B(z),$ $a$ being a real parameter, transforms (1) into 
\begin{equation}
ds^{\prime 2}=2\lambda \nu -2\mu \widetilde{\mu }
\end{equation}
where 
\begin{eqnarray}
\lambda  &=&d\sigma +2Ldz \\
\mu  &=&P^{-1}(r-i\Sigma )dz, \\
\widetilde{\mu } &=&\ P^{-1}(r+i\Sigma )d\widetilde{z}, \\
\nu  &=&dr+i\left( \Sigma _{z}dz-\Sigma _{\widetilde{z}}d\widetilde{z}%
\right) +\frac{1}{2}K\lambda .
\end{eqnarray}
and $L=\frac{ai\widetilde{z}}{P^{2}}.$

In the new coordinates, the bivectors $N$ and $M$ are given by 
\begin{eqnarray}
N_{kl} &=&\frac{r-ia}{r-i\Sigma }N_{kl}^{\prime }\,, \\
M_{kl} &=&M_{kl}^{\prime }+\frac{2PKL}{r-i\Sigma }N_{kl}^{\prime }\,,
\end{eqnarray}
where 
\begin{eqnarray}
N^{\prime } &=&\lambda \wedge \mu \\
M^{\prime } &=&\lambda \wedge \nu -\mu \wedge \widetilde{\mu }.
\end{eqnarray}

Making this transformation on the flat space line element 
\begin{eqnarray}
\underset{0}{ds}^{2} &=&2d\rho d\sigma +2kd\sigma ^{2}-\frac{2\rho ^{2}}{%
p^{2}}d\zeta d\widetilde{\zeta } \\
\underset{0}{p} &=&1+k\zeta \widetilde{\zeta },
\end{eqnarray}
writing $\overline{z}$ for $\widetilde{z}$ and putting $\sigma =u-ia\frac{z%
\overline{z}}{\underset{0}{P}}$ with $\underset{0}{P}=1+kz\overline{z}$ we
get 
\begin{equation}
ds^{2}=2\lambda (dr+i\Sigma _{z}dz-i\Sigma _{\overline{z}}d\overline{z}%
+k\lambda )-\frac{2(r^{2}+\Sigma ^{2})}{\underset{0}{P}^{2}}dzd\overline{z}
\end{equation}
where $\lambda =du+ia\underset{0}{P}^{-2}(\overline{z}dz-zd\overline{z})$
and $\Sigma =a\frac{1-kz\widetilde{z}}{1+kz\widetilde{z}}$. Note that, for $%
k=1,$ this is the transformation discussed by Newman in [5], where the
connection between the different set of coordinates is achieved via 
\begin{equation}
\zeta =\tan \frac{\theta }{2}e^{i\varphi }.
\end{equation}

\section{Linear Approximation}

We now examine perturbations of the line element (1) by writing 
\begin{equation}
p=\underset{0}{p}+\varepsilon \underset{1}{p}
\end{equation}
and working to an accuracy of the first order in $\varepsilon .$ It is
convenient to take 
\begin{equation}
\underset{1}{p}=\alpha _{\zeta }+\beta \overline{\zeta }+k\overline{\zeta }%
(\zeta \alpha _{\zeta }-2\alpha ),
\end{equation}
where $\alpha =\alpha (\sigma ,\zeta )$ and $\beta =\beta (\sigma ,\zeta )$
are arbitrary functions.

The Weyl is given by 
\begin{eqnarray}
\underset{0}{C}{}_{klmn} &=&0, \\
\underset{1}{C}{}_{klmn} &=&\frac{1}{\rho ^{2}}\underset{0}{p}\Phi
(M_{kl}N_{mn}+N_{kl}M_{mn})  \notag \\
&&-\left[ \frac{1}{\rho ^{2}}\left( \underset{0}{p}^{2}\Phi \right)
{}_{\zeta }-\frac{1}{\rho }\underset{0}{p}\left( \Psi \underset{0}{p}+\Phi
_{\sigma }\widetilde{\zeta }\right) \right] N_{kl}N_{mn}
\end{eqnarray}
\noindent where $\Phi =\beta _{\zeta \zeta }$ and $\Psi =\alpha _{\zeta
\zeta \zeta \sigma }.$

Next we transform to the coordinates $u,$ $r,$ $z,\overline{z},$ restrict
ourselves to the space-time in which $u$ and $r$ are real, $z$ and $%
\overline{z}$ are complex conjugates and we consider the line element $%
ds^{2}=\underset{0}{ds}^{2}+\varepsilon \underset{1}{ds}^{2}+\varepsilon 
\underset{1}{d\overline{s}}^{2}.$ Then, using our transformation of
bivectors and our expression for $C$ in the linear approximation we get 
\begin{equation}
C_{klmn}=\varepsilon \left[ X(M_{kl}^{\prime }N_{mn}^{\prime
}+N_{kl}^{\prime }M_{mn}^{\prime })+YN_{kl}^{\prime }N_{mn}^{\prime }\right]
+c.c.
\end{equation}
where 
\begin{eqnarray}
X &=&\frac{P}{(r-ia)(r-i\Sigma )}\Phi \,, \\
Y &=&-\frac{P}{(r-ia)^{2}}\left[ \frac{2k\overline{z}(r+ia)}{r-i\Sigma }\Phi
+P\Phi _{z}-\Psi P(r-i\Sigma )-\Phi _{\sigma }(r+ia)\overline{z}\right] .
\end{eqnarray}

\bigskip The Bel-Robinson tensor can then be written as 
\begin{eqnarray}
\frac{1}{2}P_{abcd} &=&\underset{1}{^{+}C}{}_{amnc}\underset{1}{^{-}C}%
{}_{b}{}^{mn}{}_{d}  \notag \\
&=&4\left| \Gamma \right| ^{2}\left[ \lambda _{(a}\lambda _{b}\lambda
_{c}\nu _{d)}+3\lambda _{(a}\lambda _{b}\mu _{c}\overline{\mu }_{d)}\right]
\\
&&+4\Gamma \overline{\Delta }\lambda _{(a}\lambda _{b}\lambda _{c}\overline{%
\mu }_{d)}  \notag \\
&&+4\overline{\Gamma }\Delta \lambda _{(a}\lambda _{b}\lambda _{c}\mu _{d)} 
\notag \\
&&+\left| \Delta \right| ^{2}\lambda _{a}\lambda _{b}\lambda _{c}\lambda _{d}
\notag
\end{eqnarray}
where 
\begin{eqnarray}
\Gamma &=&X\frac{r-ia}{r-i\Sigma }\,, \\
\Delta &=&\frac{2XPKL(r-ia)}{(r-i\Sigma )^{2}}+Y\left( \frac{r-ia}{r-i\Sigma 
}\right) ^{2}\,.
\end{eqnarray}

\section{An auxiliary metric and the gravitational density}

There are two ways of measuring the gravitational field: first one can look
at the differential invariants obtained from the Weyl tensor; second one
looks at the sum of the squares of its components.

We want to apply the second procedure, but we need a positive definite
metric for that. We can create one and take the sum of the squares using
this auxiliary metric.

Let 
\begin{equation}
\gamma _{ab}=2t_{a}t_{b}-t_{r}t^{r}g_{ab}
\end{equation}
where $t$ is a timelike (unit) vector. We have [3] 
\begin{equation}
^{+}C_{abcd}{}^{-}C_{rstu}\gamma ^{ar}\gamma ^{bs}\gamma ^{ct}\gamma
^{du}=\left( t_{r}t^{r}\right) ^{2}P_{abcd}t^{a}t^{b}t^{c}t^{d}.
\end{equation}
This expression usually depends on $t$ and is of no interest, but in the
linear approximation we can take $t$ to be constant with respect to the
background (Kerr metric in this case). The easiest way to find such a vector
field is to go to Cartesian coordinates. This is accomplished by the
following transformation (see [4]) 
\begin{eqnarray}
U &=&u+\frac{rz\overline{z}}{P} \\
V &=&\frac{r}{P}+ku \\
Z &=&\frac{r-ai}{P}z
\end{eqnarray}
where the capital letters represent the Minkowski coordinates. Any constant
timelike one form, i.e. 
\begin{equation}
c_{0}dU+c_{1}dV+c_{2}dZ+\overline{c}_{2}d\overline{Z}
\end{equation}
$c_{i}=const.(i=1,2,3)$ such that $c_{0}c_{1}-c_{2}\overline{c}_{2}>0,$
represents a constant timelike one-form in Kerr coordinates.

\bigskip

\section{Discussion}

\bigskip

\noindent (a) For $k=0$ a constant tetrad is 
\begin{eqnarray}
\widehat{\nu } &=&\nu \\
\widehat{\mu } &=&\mu +z\nu \\
\widehat{\lambda } &=&\lambda +\overline{z}\mu +z\overline{\mu }+z\overline{z%
}\nu .
\end{eqnarray}
A convenient timelike one form is $\tau =\widehat{\lambda }+\widehat{\nu };$
let $t$ be the vector field associated with it. We have 
\begin{equation}
\frac{1}{2}P_{abcd}t^{a}t^{b}t^{c}t^{d}=\left\{ 4\left| \Gamma \right|
^{2}+\left| 4\overline{z}\Gamma -(1+z\overline{z})\Delta \right|
^{2}\right\} (1+z\overline{z})^{2},
\end{equation}
and 
\begin{eqnarray}
\Gamma &=&\frac{\Phi }{\left( r-ia\right) ^{2}}, \\
\Delta &=&Y|_{k=0}  \notag \\
&=&-\frac{1}{(r-ia)^{2}}\left[ \Phi _{z}-\Psi (r-ia)-\Phi _{\sigma }(r+ia)%
\overline{z}\right] .
\end{eqnarray}
One can get directional nonsingular and asymptotically flat solutions by
setting $\Phi =\frac{1}{\left( \sigma -i\right) ^{n}}$ with $n>1$ and $\Psi =%
\frac{1}{\left( \sigma -i\right) ^{m}}$ with $m>1.$

(b) If $k>0$, a constant timelike one form is $\tau =k\lambda +\nu .$ We
have 
\begin{equation}
\frac{1}{2}P_{abcd}t^{a}t^{b}t^{c}t^{d}=4k\left| \Gamma \right| ^{2}+\left|
\Delta \right| ^{2}.
\end{equation}

For $a=0$ we have 
\begin{eqnarray}
\Gamma &=&\frac{P}{r^{2}}\Phi , \\
\Delta &=&Y|_{a=0}  \notag \\
&=&-\frac{P}{r^{2}}\left[ 2k\overline{z}\Phi +P\Phi _{z}-\Psi Pr-\Phi
_{\sigma }r\overline{z}\right]
\end{eqnarray}
The situation is similar to the null case, directional singularities being
unavoidable.

\bigskip

\bigskip

\noindent \textbf{References}

\noindent \lbrack 1] Trautman A (1962) ''Analytic solutions of
Lorentz-invariant linear equations'' \emph{Proc. Roy. Soc. Ser. A} \textbf{%
270} 326.

\noindent \lbrack 2] Synge J L (1956)\ ''Relativity: the special theory''
North-Holland Publishing Co., Amsterdam; Interscience Publishers, Inc., New
York; 1956.

\noindent \lbrack 3] Robinson I (1997) ''On the Bel-Robinson tensor.
Geometry and physics'' \emph{Classical Quantum Gravity} \textbf{14}, no. 1A,
A331.

\noindent \lbrack 4] Robinson I, Robinson J Zund J D (1969) ''Degenerate
gravitational fields with twisting rays'' \emph{J. Math. Mech}. \textbf{18}
881.

\noindent \lbrack 5] Newman E T (1973) ''Complex coordinate transformations
and the Schwarzschild-Kerr metrics'' \emph{J. Math. Phys.} \textbf{14}, no.
6, 774.

\end{document}